\documentclass{aa}  
\usepackage{float}
\usepackage{caption}
\usepackage{subcaption}
\usepackage{graphicx}
\usepackage{txfonts}
\usepackage{hyperref}
\begin{document}

   \title{Confirmation Of Two New Galactic Bulge Globular Clusters: FSR 19 and FSR 25}


   \author{C.Obasi
          \inst{1,2}
          \and
          M. G\'omez\inst{1}
           \and
          D.Minniti\inst{1,3}
           \and
          J.Alonso-Garc\'ia\inst{4,5}
          }

   \institute{Depto. de Ciencias F\'isicas, Facultad de Ciencias Exactas, Universidad Andres Bello,\\ Av. Fernandez Concha 700, Las Condes, Santiago, Chile.
              \email{c.obasi@uandresbello.edu}
         \and
             Centre for Basic Space Science, University of Nigeria Nsukka\\
                     \and
             Vatican Observatory, V00120 Vatican City State, Italy.\\
                     \and
             Centro de Astronom\'{i}a (CITEVA), Universidad de Antofagasta, Av. Angamos 601, Antofagasta, Chile.
                      \and
                      Millennium Institute of Astrophysics, Nuncio Monse\~nor Sotero Sanz 100, Of. 104, Providencia, Santiago, Chile.
             }

   \date{Accepted }

 
  \abstract
   {
Globular clusters (GCs) in the Milky Way (MW) bulge are very difficult to study because: i) they suffer  from the severe crowding and galactic extinction; which are characteristic of these inner Galactic regions  ii) they are more prone to be affected by dynamical processes. Therefore, they are relatively faint and difficult to map. However, deep near-infrared photometry like that provided by the VISTA Variables in the Via Láctea Extended Survey (VVVX) is allowing us to map GCs in this crucial yet relatively uncharted region.
}
   {Our main goals are to study the true nature of the GC candidates FSR 19 and FSR 25 and measure their physical parameters.}
   {We use the near-infrared (NIR) VVVX database, in combination with the Two Micron All Sky Survey (2MASS) and Gaia EDR3 proper motions (PMs) and photometry to study ages, metallicities, distances, reddening, mean PMs, sizes and integrated luminosities for FSR 19 and FSR 25. A robust combination of  selection criteria
  allows us to effectively clean interlopers among our samples.
}
   {Our results confirm with high confidence that both FSR 19 and FSR 25  are genuine MW bulge GCs. Each of the performed tests and resulting parameters provides clear evidence for the GC nature of these targets. We derive distances of 7.2$\pm$0.7 kpc and  D=7.0$\pm$0.6 (corresponding to distance moduli of 14.29$\pm$0.08 and 
   14.23$\pm$0.07) for FSR 19 and FSR 25, respectively. Their ages and metallicities are 11 Gyr and [Fe/H]= -0.5 dex for both clusters, which were determined from Dartmouth and   PARSEC isochrone fitting. The integrated luminosities are M$_{Ks}$(FSR 19) = -7.72 mag and  M$_{Ks}$(FSR 25) = -7.31 mag which places them in the faint tail of the GC Luminosity Function. By adopting a King profile for their number distribution, we  determine their core and tidal radii ($r_c$, $r_t$). For FSR 19, r$_{c}$= 2.76$\pm$0.36 pc and  r$_{t}$=5.31$\pm$0.49 pc, while FSR 25 appears more extended with r$_{c}$= 1.92$\pm$0.59 pc and  r$_{t}$=6.85$\pm$1.78 pc. Finally their mean GC PMs (from Gaia EDR3) are  $\mu_{\alpha^\ast}$= -2.50 $\pm$0.76 mas $yr^{-1}$, $\mu_{\delta}$= -5.02 $\pm$0.47 mas $yr^{-1}$ and $\mu_{\alpha^\ast}$= -2.61 $\pm$ 1.27 mas $yr^{-1}$ , $\mu_{\delta}$= -5.23 $\pm$0.74 mas $yr^{-1}$  for  FSR 19 and FSR 25, respectively. }
   {
We have demonstrated and confirmed based on the measured astrophysical parameters that the two target clusters are indeed genuine and of low luminosity relatively metal-rich old GCs in the bulge of the MW.
}


   \maketitle
%

\section{Introduction}\label{intro}

The confirmation of each new Globular Cluster (GC) of the Milky Way (MW) is a treasure, as GCs are important tracers of the field stellar populations, and provide valuable evidence for the formation of the Galaxy. Metal-poor GCs located in the Galactic bulge in particular may probably be the oldest objects in the Galaxy \citep[e.g.,][]{barbuy2016high}.

The new catalogue of Star Clusters, Associations and Candidates in the Milky Way by \cite{bica2019vizier} contains about 10,000 star clusters, including many new low-luminosity candidates discovered by the VVV survey \citep[e.g.,][]{minniti2017new,camargo2018five,garro2020vvvx,
garro2021confirmation}, significantly increasing the bulge GC sample. While many of these have already been confirmed or discarded by follow up studies \citep[e.g.,][]{Gran,palma2019analysis,piatti2018physical}, it is clear that the bulge GC census is incomplete \citep{ivanov2005red,minniti2017fsra}.

In this work we use the  VISTA Variables in the Via Láctea Extended Survey (VVVX), Two Micron All Sky Survey (2MASS) and Gaia EDR3 photometry Survey data to study the properties of two MW bulge GC candidates. These are FSR 19 and FSR 25, first discovered by \citet*{froebrich2007systematic} and located at $(l, b) = 5.499, 6.071$ deg and $7.534, 5.649$ deg, respectively. 
\citet*{froebrich2007systematic} searched for new star clusters in the MW using 2MASS, \cite[][]{skrutskie2006two}, presenting 1021 new candidates, FSR 19 and FSR 25 among them. \cite{buckner2013properties} developed an automatic method to estimate distances and reddenings to 378 previously known open clusters and 397 new cluster candidates from \citet*{froebrich2007systematic}. This technique was based on JHK photometry alone 
and relied on models for the foreground contamination. The estimated distances to FSR 19 and FSR 25 were 3.80 kpc and 3.60 kpc respectively, therefore presumably away from the MW bulge. \cite{kharchenko2013global} combined kinematic and photometric membership to determine physical parameters for a large sample of star clusters in a homogeneous way. 

\begin{figure*}
   \centering
   \includegraphics[width=10cm, height=12.94cm]{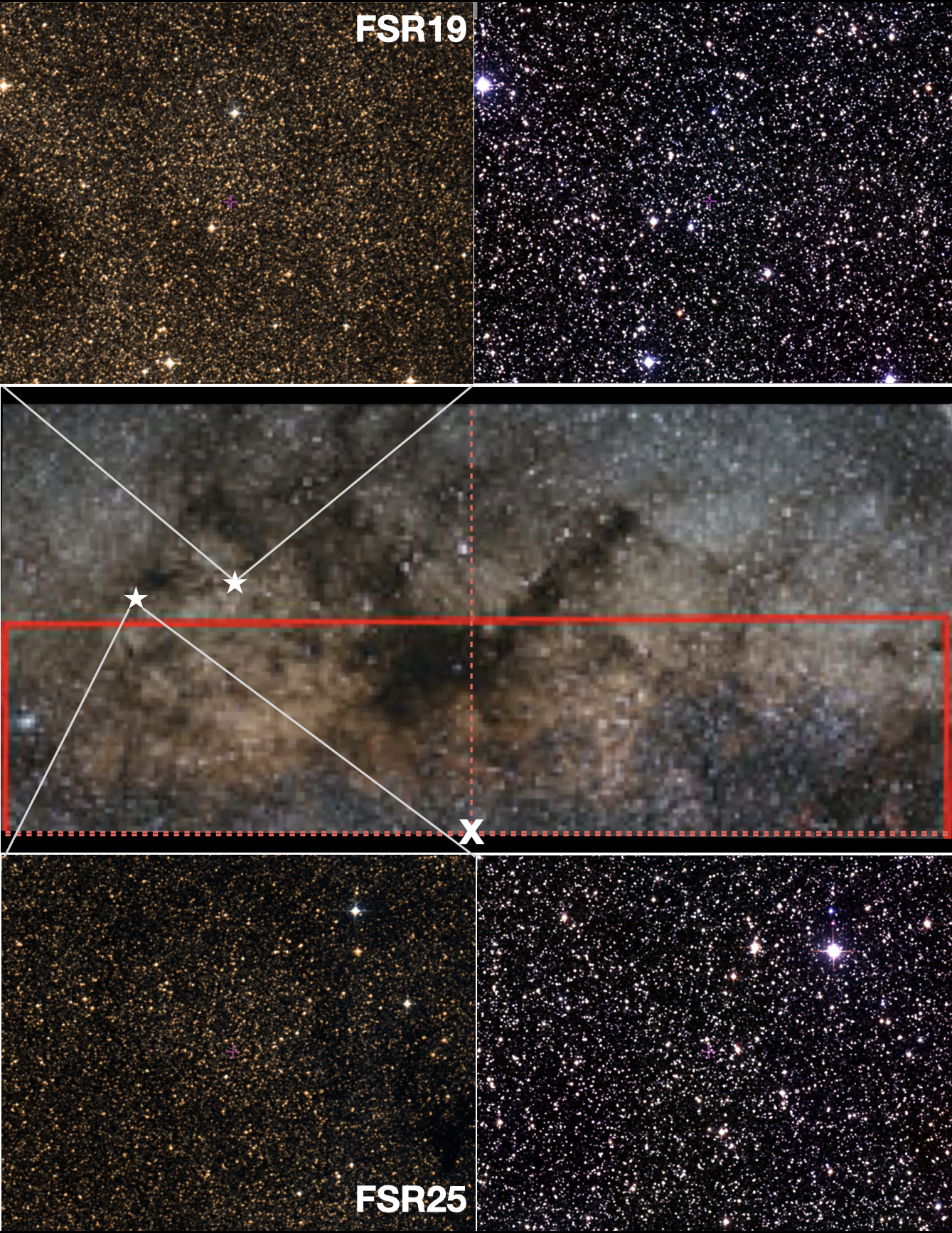}
   \caption{A zoom-in image of the clusters in both optical (left) and near
IR (right) for FSR 19 (upper frame) and FSR 25 (bottom frame). The
middle frame shows the complex bulge region where the clusters are
located. The position of the Galactic centre is shown with a white X. The
red rectangle is the region covered by the VVV Survey, and the top part corresponds to the extension VVVX. The total field of view
shown is 20X10 sqdeg. The Galactic major and minor axes are shown
with dotted red lines.}
              \label{Fig1}%
    \end{figure*}

In particular, they obtained D = 5.98 kpc and $M_{Ks}$=-7.96 mag for FSR 19, and D=6.64 kpc and $M_{Ks}$= -8.80 mag for FSR 25 respectively. 
Fig. \ref{Fig1} shows that this is a complicated region in terms of heavy and non-uniform extinction. Just to illustrate this, the colour excess in the Gaia bands  (BP-RP) changes from E(BP-RP) = 0.2 mag to 1.4 mag in a scale of < 2 arcmin for FSR 25 and slightly less for FSR 19.  However, we use VVVX and 2MASS near-infrared (NIR) photometry, which is less sensitive to extinction. 
A summary of the basic parameters derived for these clusters in this work are listed in table \ref{table:1}.

In this paper we confirm FSR 19 and FSR 25 as genuine low-luminosity members of the Galactic bulge GC system, measuring their parameters and estimating their ages based on NIR data from the 2MASS and VVVX Survey, combined with Gaia EDR3 optical data. In section \ref{sec:Data}  a description of the dataset is given. Section \ref{sec:phyical} describes the physical characterization of the derived parameters such as reddening, extinction, color-magnitude diagrams (CMDS), ages, metallicities, luminosities and radii. In the light of our findings, we discuss their physical nature in section \ref{sec:Dis} and finally summarize our conclusions in section \ref{sec:concl}.

\section{Gaia EDR3, 2MASS and VVVX data.}\label{sec:Data}
Non-uniform and severe extinction towards the bulge of our Galaxy  \citep[e.g.,][]{schlafly2011measuringData,alonso2017extinction,alonso2018milkyData} is undoubtedly a major issue in photometric studies. Although far less than in the optical wavebands, NIR observations do suffer from this effect. From the already completed VVV Survey, exquisite extinction maps have been derived \citep{alonso2017extinction,
surot2020mapping} which allow for a very detailed and homogeneous correction.

The VVVX Survey \citep{minniti2018mapping} maps the Galactic bulge and southern disk in the NIR with the VIRCAM (VISTA InfraRed CAMera) at the 4.1 m wide-field Visible and Infrared Survey Telescope for Astronomy \citep[VISTA;][]{emerson2010visibleData} at  the European Southern Observatory (ESO) Paranal Observatory (Chile). In the Galactic bulge, the VVVX Survey covers about 600 sqdeg., using the J  (1.25 $\mu$m), H  (1.64 $\mu$m), and $K_s$  (2.14 $\mu$m) NIR passbands. The VVVX Survey data reduction and the archival merging were carried out at the Cambridge Astronomical Survey Unit \citep{ irwin2004vistaData}(CASU) and VISTA Science Archive (VSA) at the Wide-Field Astronomy Unit (WFAU), using the VISTA Data Flow System \citep{cross2012vista}. In order to deal with the high crowding in this region, we extracted the the point spread function (PSF) photomety using the pipeline described in \cite{alonso2018milkyData}.

2MASS is an all sky survey in the NIR bands J(1.25 $\mu$m), H(1.65 $\mu$m), and $K_s$(2.16 $\mu$m)\citep{cutri2003explanatory,skrutskie2006two}  observed with two dedicated telescopes located in both hemispheres working synchronously. The integration time of 7.8 s reaches  15.8, 15.1 and 14.3 mag at the J, H, and $K_s$  bands respectively. The data have been processed by 2MASS production Pipe-line system 2MAPPS \citep{skrutskie2006two}.

 Gaia EDR3 contains the apparent brightness in G magnitude for over $1.8\times10^9$ sources which are brighter than 21 mag, and for $1.5\times10^9$ sources, passbands $G_{BP}$ covering 330-680 nm and $G_{RP}$ covering 630-1050 nm, which are not available in DR1 and DR2. PMs are available for $1.4\times10^9$ sources with an accuracy of 0.02 mas $yr^{-1}$ for sources brighter than G=15 mag, 0.07 mas $yr^{-1}$ for sources brighter than G=17 mag and 0.5 mas $yr^{-1}$ for sources brighter than G=20 mag \citep{gaia2020}. The data have been processed by the Gaia Data Processing and Analysis Consortium (DPAC).
 
 We have used a combination of the NIR  VVVX Survey together with 2MASS and optical Gaia EDR3 data. As we show in Section \ref{sec:phyical}, this robust combination opens new windows for studies of highly reddened and crowded regions like the MW bulge.

\begin{figure*}%
\centering
\begin{subfigure}{.8\columnwidth}
\includegraphics[width=\columnwidth]{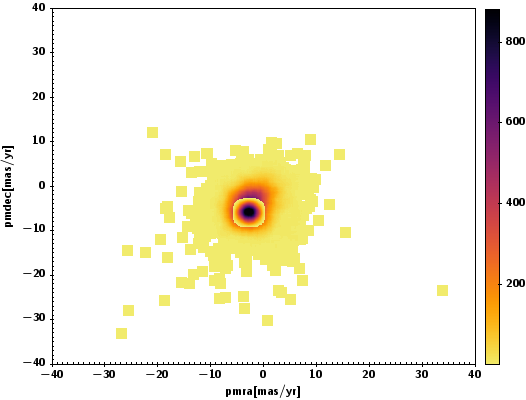}%
\label{sub:vpa}%
\end{subfigure}\hfill%
\begin{subfigure}{.8\columnwidth}
\includegraphics[width=\columnwidth]{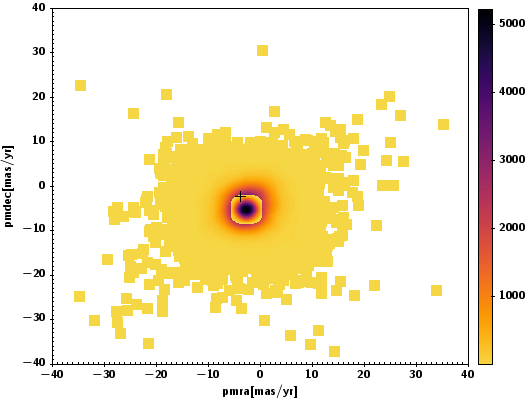}%
\label{sub:vpb}%
\end{subfigure}\hfill%
\begin{subfigure}{.8\columnwidth}
\includegraphics[width=\columnwidth]{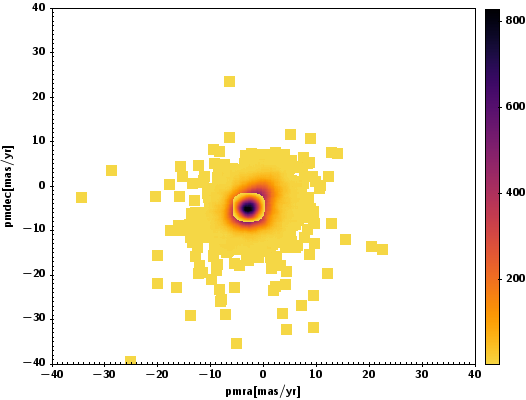}%
\label{sub:vpc}%
\end{subfigure}\hfill%
\begin{subfigure}{.8\columnwidth}
\includegraphics[width=\columnwidth]{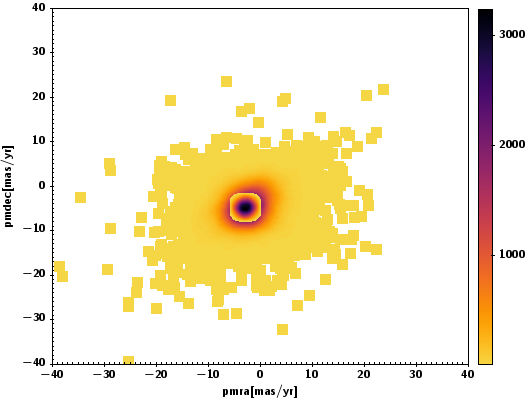}%
\label{sub:vpd}%
\end{subfigure}%
\caption{Vector PM diagrams for the bright sources (Ks<12 mag) matched with the 2MASS+Gaia catalogues and the fainter sources (Ks>12 mag) matched with VVVX+Gaia catalogues(upper frame FSR 19 and bottom frame FSR 25). The yellow region shows the cluster selection and the black cross in (upper right frame) represents the RR Lyrae found within 5$'$ of the FSR 19 cluster. The color bars indicates the concentration level.}
\label{fig:vpm}
\end{figure*}

We distilled a  sample of relatively contaminant-free  catalogue of most probable clusters members, drawn from the precise astrometry and PMs from Gaia EDR3, and also combining 2MASS+Gaia and VVVX+Gaia catalogues in a way to incorporate both bright as well as the fainter sources. Stars with Ks$<$11 mag are saturated in VVVX photometry 
and we discarded all nearby stars with a parallax>0.5 mas. Consequently, we scrutinised the vector PM (VPM) diagrams in Fig. \ref{fig:vpm} which shows a sharp peak with respect to the stellar background distributions,  which we ascribed as the clusters mean PMs. 
 These mean PMs as measured by  Gaia EDR3 are as $\mu_{\alpha^\ast}$= -2.50 $\pm$0.76 mas $yr^{-1}$, $\mu_{\alpha^\ast}$= -2.61 $\pm$ 1.27 mas $yr^{-1}$  and  $\mu_{\delta}$= -5.02 $\pm$0.47 mas $yr^{-1}$ , $\mu_{\delta}$= -5.23 $\pm$0.74 mas $yr^{-1}$  for  FSR 19 and FSR 25 with respect to the mean clusters PMs. We showed the VPM diagrams for brighter sources with Ks$<$12 mag matched in the 2MASS+Gaia catalogues and the fainter sources with Ks$>$12 matched in the VVVX+Gaia catalogues in Fig. \ref{fig:vpm} (upper frame),  for FSR 19 and Fig. \ref{fig:vpm} (bottom frame), for FSR 25. For the rest of our analysis, we selected stars within  2.0 mas $yr^{-1}$  of these mean VPM values (after extensive testing of the selection criteria :1.0, 1.5, 2.0, 2.5 mas $yr^{-1}$ ).  Stars outside of the 2.0 mas $yr^{-1}$  circle selected on the VPM diagrams but  within the selected radius of 1.2$'$ of the clusters centre, shown in yellow in Fig. \ref{fig:vpm}, are likely stars from the surrounding field. We did not consider them in the following analysis.

\section{Physical characterization of the GC candidates FSR 19 and FSR 25.} \label{sec:phyical}
\begin{figure*}%
\centering
\begin{subfigure}{.8\columnwidth}
\includegraphics[width=\columnwidth]{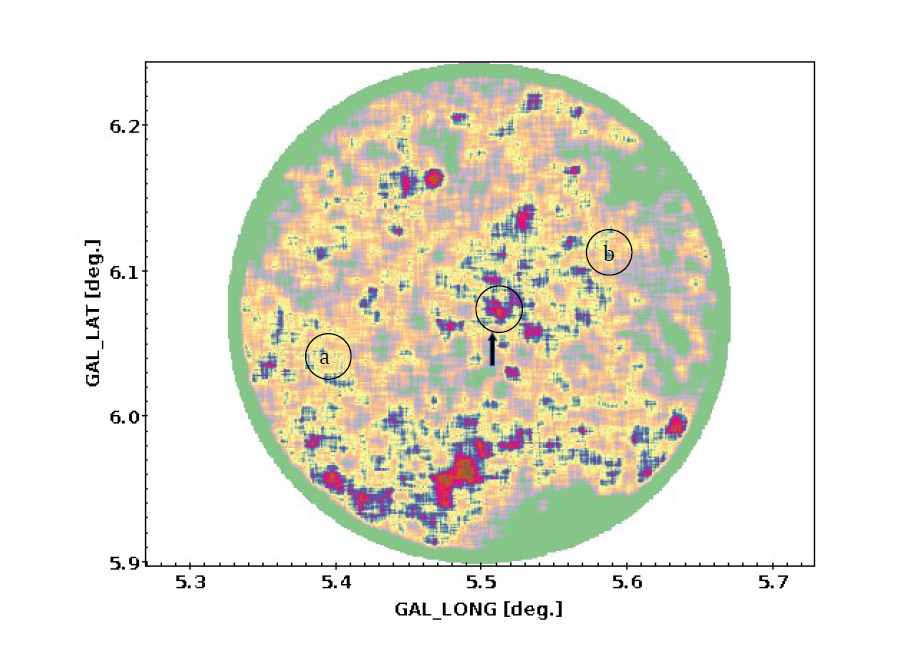}%
\label{sub:gaiadensitya}%
\end{subfigure}\hfill%
\begin{subfigure}{.8\columnwidth}
\includegraphics[width=\columnwidth]{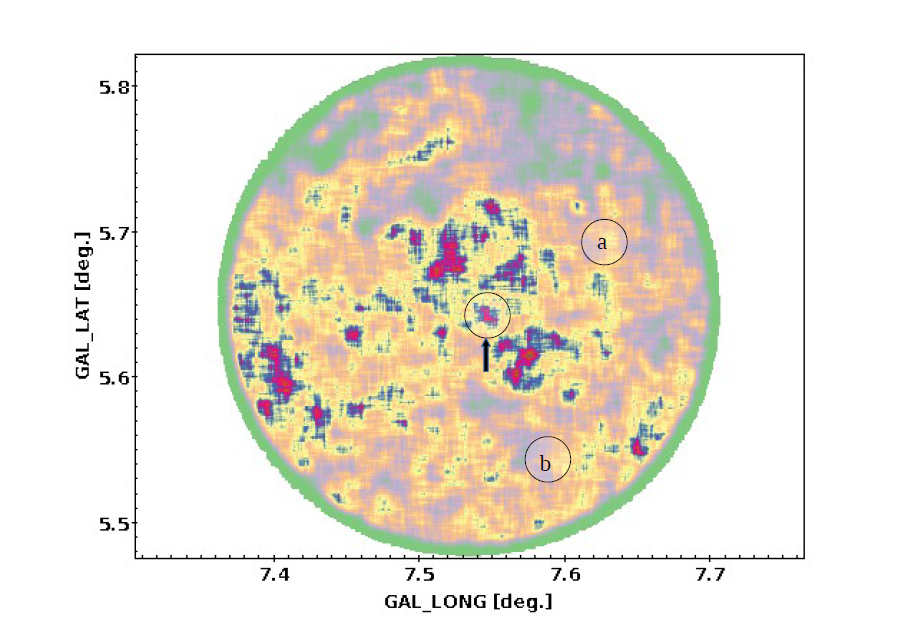}%
\label{sub:gaiadensityb}%
\end{subfigure}%
\caption{Gaia stellar density map for  r=1.2 $'$ are shown in the Galactic coordinates indicating the positions of the two clusters (FSR 19 left and FSR 25 right) with an arrow. The comparison fields are also represented with a and b.}
\label{fig:gaia}
\end{figure*}

Both clusters are projected next to complex filamentary dark nebulae (FSR 19 in the vicinity of Barnard 268 and FSR 25 next to Barnard 276).
Therefore it is important to consider first the possibility that they could be merely background fluctuations. 
Fig. \ref{fig:gaia} shows the optical Gaia source density maps of these regions. Clearly, an excess of stars above the background value is found at the location of FSR 19 and FSR 25. We select member stars within a circle of r= $\sim$1.2$'$ centred at  the coordinates 
($\alpha, \delta)$ = (263.915, -21.061) deg. and (265.437,-19.596) deg. for FSR 19 and FSR 25 respectively. These correspond to $(l,b)$ = (5.499,6.071) deg. and (7.534,5.649) deg.

We compared the stellar densities of these central regions with those of randomly selected backgrounds fields of equal radii,  which are labeled as "a" and "b" in Fig. \ref{fig:gaia}. In support of the GC nature of the candidates, the luminosity functions (LFs) of the cluster regions are very different from those of the background fields. 
We tested other control fields as well and in all cases there is a clear peak of red clump (RC) stars for the cluster regions which is never seen in the background fields. Overdensities caused by the presence of RC stars are well studied features of bulge GCs and a strong evidence for their nature. RC stars are also tracers of prominent structures in the inner regions of the Galaxy \citep{gonzalez2015reinforcing}, in  our case, our kinematic selection eliminates most field RC giants.

\begin{figure*}
   \centering
   \includegraphics[width=18cm, height=10cm]{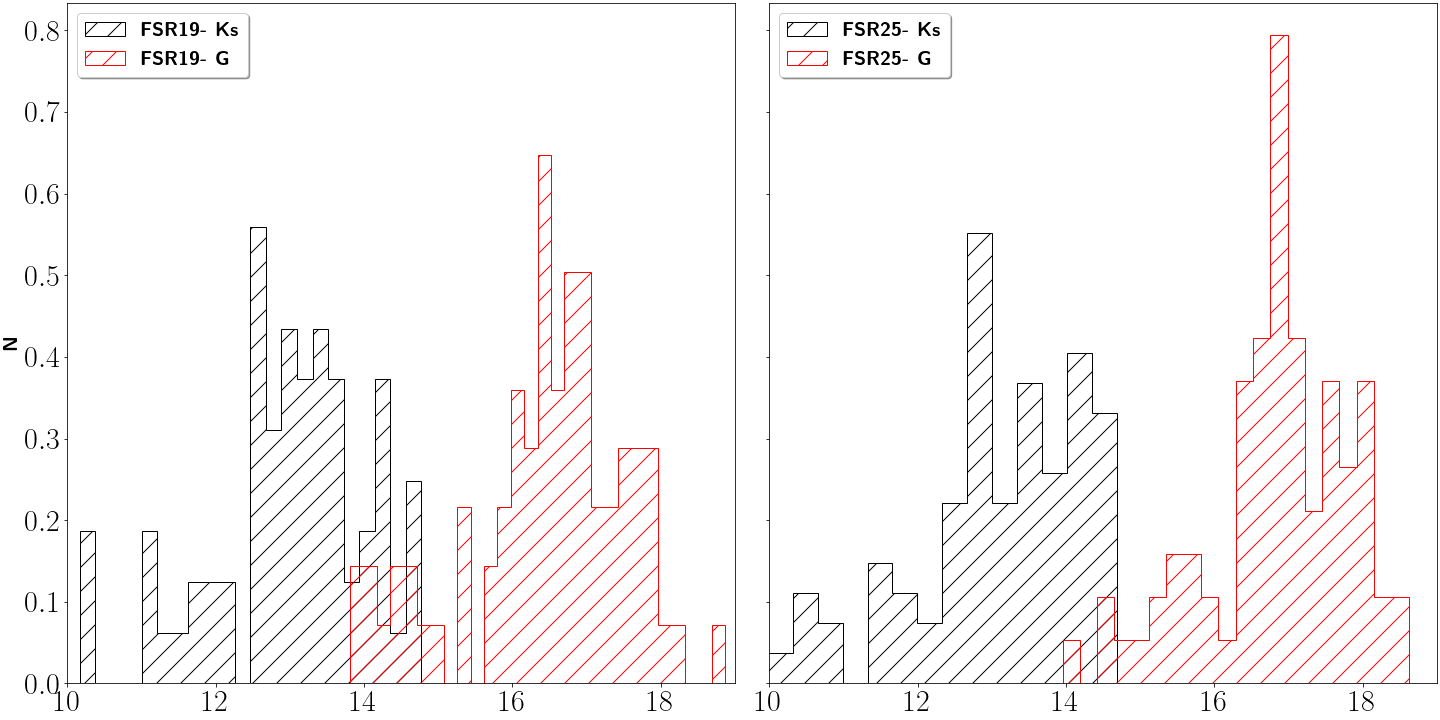}
   \caption{The luminosity functions of both FSR 19 and FSR 25 in the near IR Ks-band photometry in black and the Gaia EDR3 G-band photometry in red. In both the left and right panel we show only sources selected according to PM. The RC positions are clearly defined.   }

              \label{Fig4}%
    \end{figure*}

Fig. \ref{Fig4} shows the luminosity functions in the G and Ks bands for a 1.2$'$  radius field centred on the clusters for all objects with J-Ks > 0.7 mag 
for the PM-selected red giants members. 
These LFs clearly reveal the peaks due to the clusters RC giants at Ks=12.85$\pm$0.05, 12.88$\pm$0.06 and G=17.10$\pm$0.01 and G=17.30$\pm$0.02 for FSR 19 and FSR 25.

\begin{figure*}
   \centering
   \includegraphics[width=18cm, height=10cm]{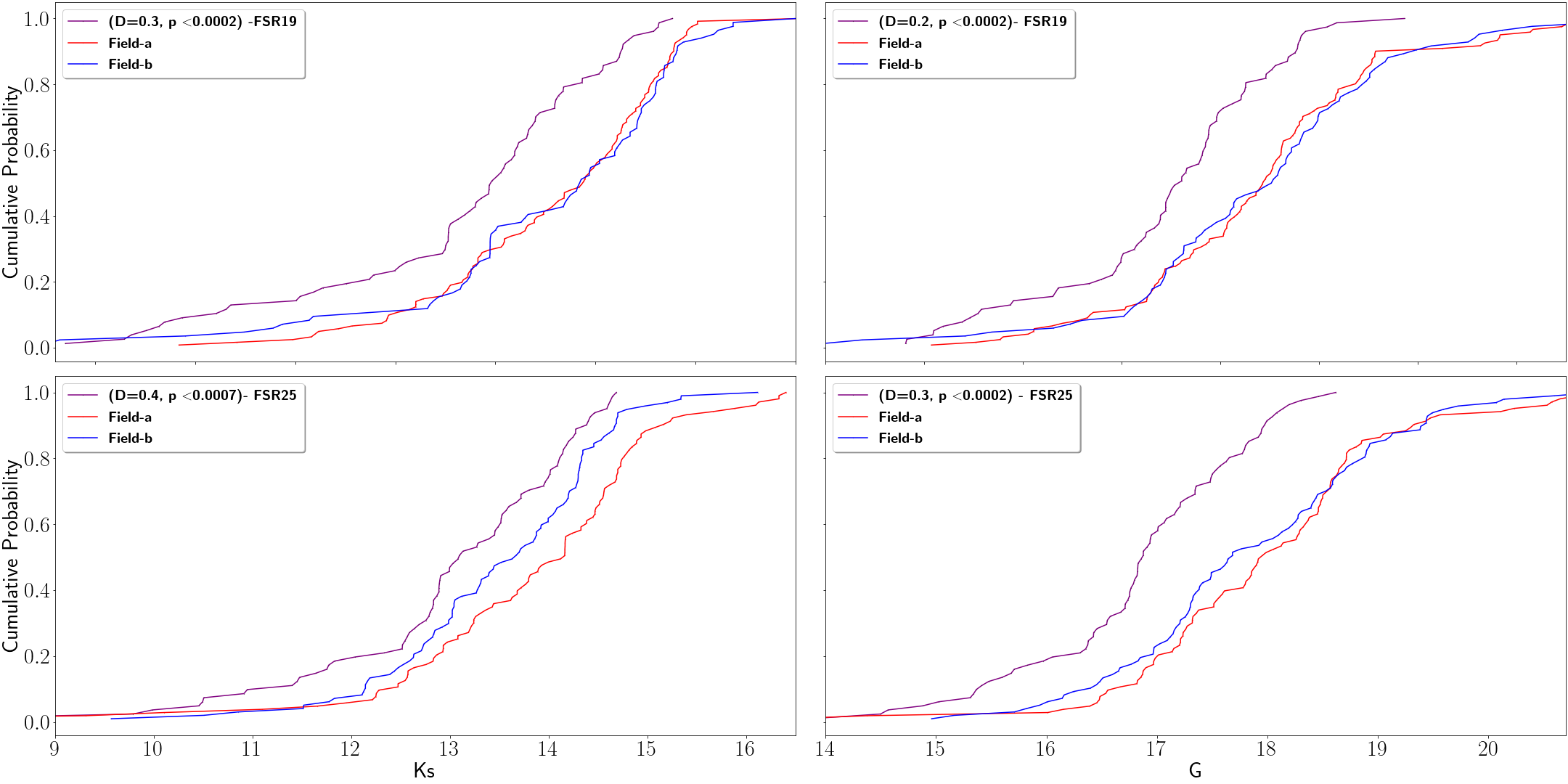}
   \caption{The cumulative probability of both targets clusters are shown with their respective comparison fields. For FSR 19 the fluctuations in the background comparison fields seem fairly uniform and varies in the field of FSR 25. The D statistic and calculated p-value are shown in the legend of each panel.}
              \label{Fig5}%
    \end{figure*}

A well established indicator to estimate the probability of two samples being drawn from the same parent distribution is the Kolmogorov-Smirnov (KS) test. The null hypothesis (no difference in the origin of both samples) means for our study that cluster and background field are statistically indistinguishable. The alternative hypothesis implies that the two LFs represent samples with different origins. The cumulative LF for each of the clusters with their respective comparison fields is shown in Fig. \ref{Fig5} The results of the KS test favoured the alternative hypothesis with a p-value of <0.0001 for both Ks and G mag LFs. Again, this points to a genuine GC nature of these targets.

\begin{figure*}
   \centering
   \includegraphics[width=15cm, height=20cm]{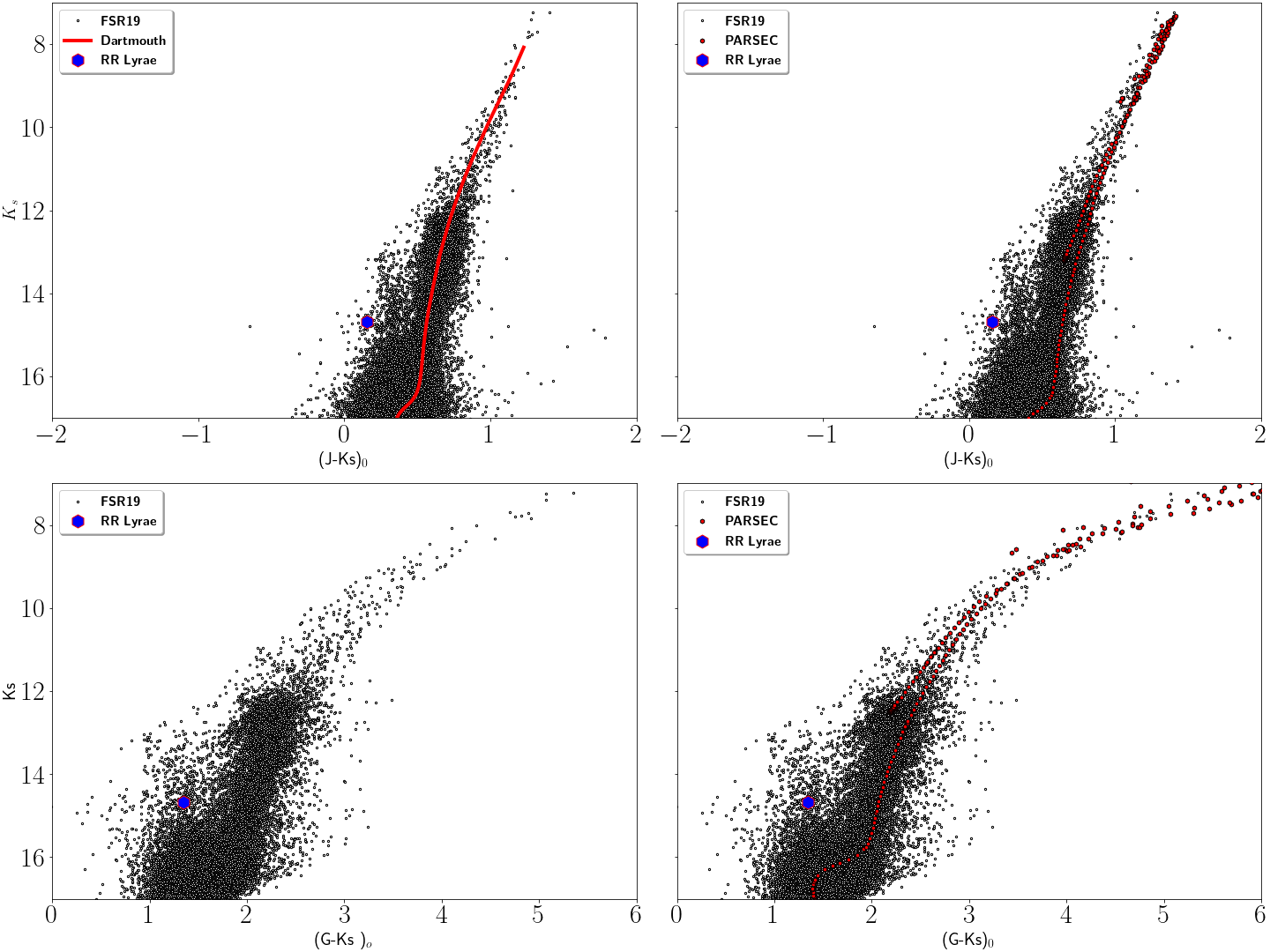}
   \caption{Optical and NIR PM decontaminated CMDs for FSR 19 cluster fitted with (Dartmouth isochrone left panel and PARSEC isochrone right panel). In all cases, the RGB stars are narrow, fainter and redder than the fields. The position of RR Lyrae found in the cluster is indicated with the blue symbol.}
              \label{Fig6}%
    \end{figure*}

\begin{figure*}
   \centering
   \includegraphics[width=15cm, height=20cm]{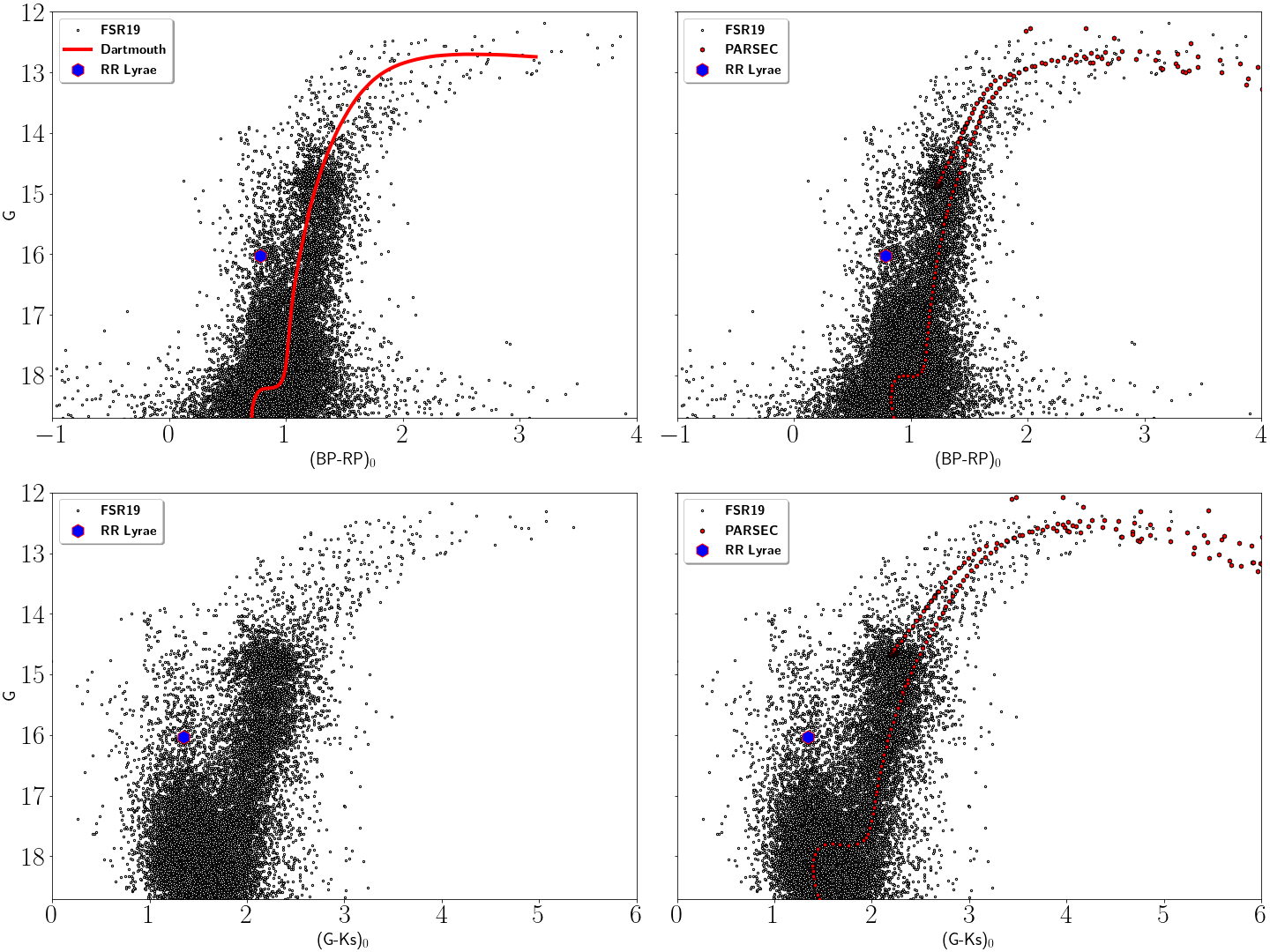}
   \caption{Optical Gaia and NIR  PM decontaminated CMDs for FSR 19 fitted with (Dartmouth left panel and PARSEC  right panel) isochrone. The RGB stars are narrow, fainter and redder than the fields.\textbf{ The position of RR Lyrae found in the cluster is indicated with the blue symbol.}}
              \label{Fig7}%
    \end{figure*}

\begin{figure*}
   \centering
   \includegraphics[width=15cm, height=20cm]{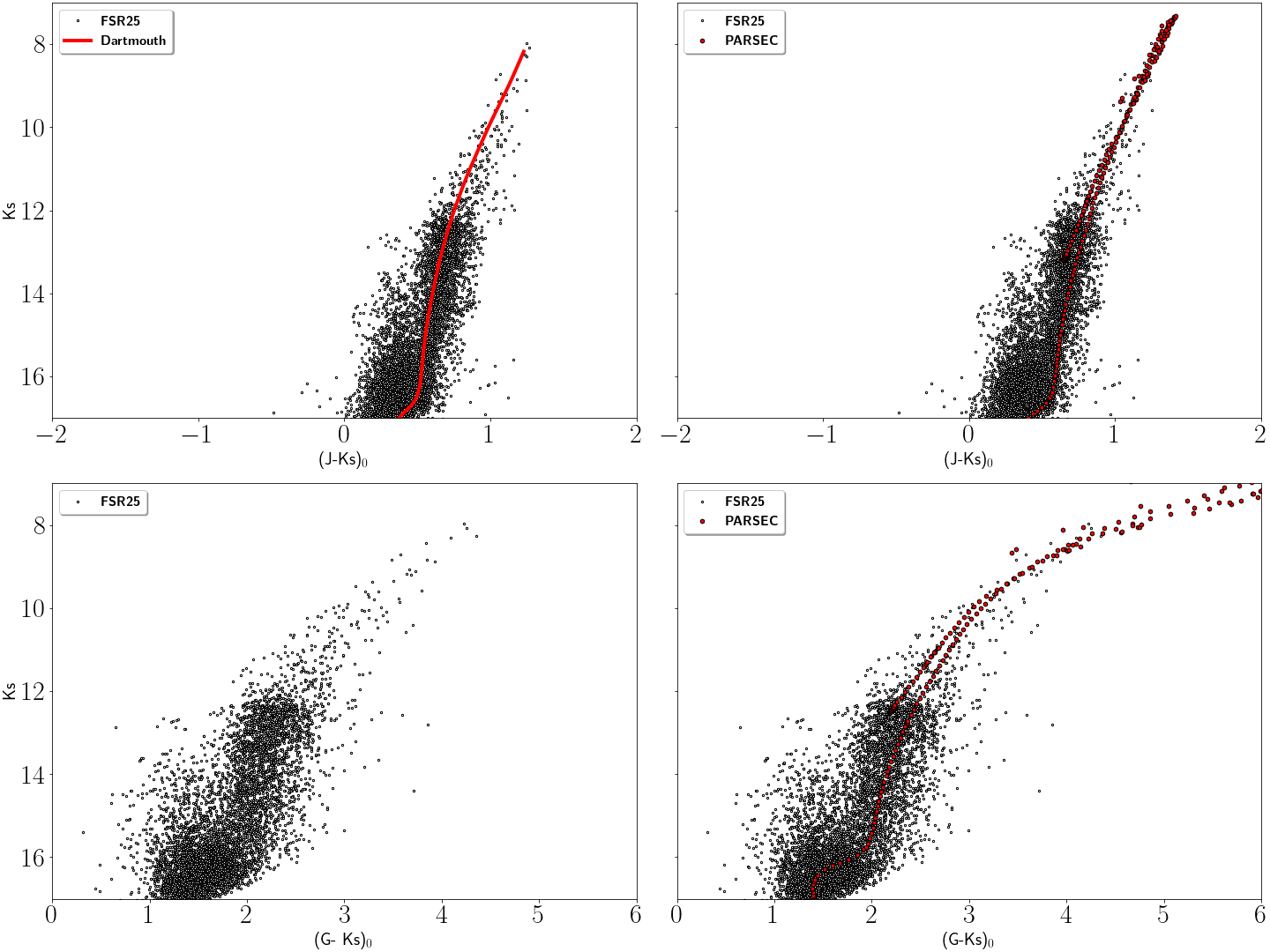}
   \caption{NIR and optical Gaia PM decontaminated CMDs for FSR 25 fitted with (Dartmouth left panel PARSEC  right panel) isochrone. The RGB stars are narrow, fainter and redder than the fields.}
              \label{Fig8}%
    \end{figure*}

\begin{figure*}
   \centering
   \includegraphics[width=15cm, height=20cm]{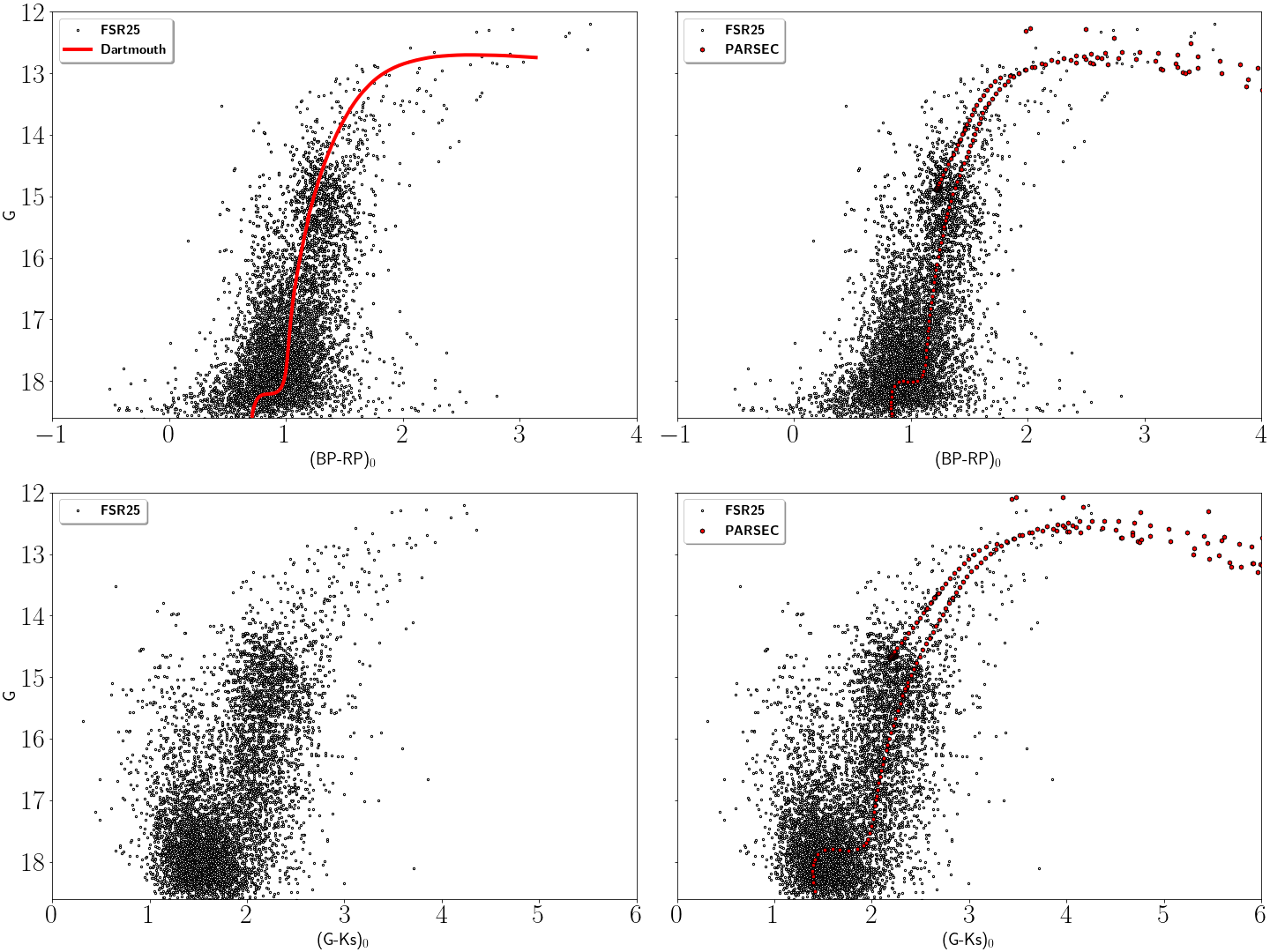}
   \caption{Optical Gaia and NIR PM decontaminated CMDs for FSR 25 cluster fitted with (Dartmouth left panel and PARSEC right panel) isochrone. In all cases, the RGB stars are narrow, fainter and redder than the fields.}
              \label{Fig9}%
    \end{figure*}

In Figs. \ref{Fig6}, \ref{Fig7}, \ref{Fig8}  \ref{Fig9} we show the color-magnitude diagram (CMD) of the clusters members that were selected from their PMs. The red giant branches (RGBs) are narrower, redder and fainter than what is observed for the neighbouring fields. This supports the argument that the field stars are mainly coming from the foreground, less reddened populations in the Galactic plane  \citep{garro2020vvvx} and we can hereto rule out any form of low extinction window which could appeared disguised as stellar overdensities  \citep[e.g.][]{minniti2018new,saito2020vvv}. Indeed, there remain some contaminating sources with colors (J-Ks) $\sim$ 0.2 mag and (BP-RP)$\sim$0.5 mag, that correspond to stars from the distant disk or the bar, or even the extended cluster horizontal branches.

We corrected for extinction following the maps of  \cite{gonzalez2012reddeningData,surot2020mapping} in the NIR and \cite{schlafly2011measuringData} in optical wavebands. 

We used the RC calibration of \cite{ruiz2018empirical} where the absolute magnitude in the Ks-band is M$_{Ks}$= (1.601$\pm$0.009) and the mean color (J-Ks)$_{0}$= (0.66$\pm$0.02). The comparison with our previously derived values for the RC  
magnitudes and colours result in the colour excesses and absorptions E(J-Ks)=(0.38$\pm$0.06), (0.54$\pm$0.09)  and A$_{Ks}$=(0.19$\pm$0.07) , (0.27$\pm$0.01) mag for both FSR 19 and FSR 25 respectively.

Distance moduli can then be estimated by adopting A$_{Ks}$/E(J-Ks)=0.5 
which yields (m-M)$_0$=(14.29$\pm$0.08) mag and (14.23$\pm$0.07) mag for both FSR 19 and FSR 25. This translates into distances to the clusters 
D=(7.2$\pm$0.7) kpc and (7.0$\pm$0.6) kpc for FSR 19 and FSR 25.

In order to test the robustness of our result, we have used the theoretical RC magnitude calibrations from \cite{alves2002k} M$_{Ks}$=-1.60 mag  and \cite{salaris2002population} M$_{Ks}$=-1.54 mag.

The results were in excellent agreement (m-M)$_0$=(14.29) mag,(14.23) mag and D=7.2 kpc,7.0 kpc for FSR 19 and (m-M)$_0$= (14.23) mag, (14.17) mag, D=7.0 kpc, 6.8 kpc for FSR 25.

In addition, we also double checked our parameters using Gaia EDR3 optical photometry. Adopting the  extinction values previously derived (A$_{Ks}$= 0.19 
and 0.27 
for FSR 19 and FSR 25), these translates into Gaia G-band  A$_G$=1.49  and 2.10 , respectively. Assuming a V-band extinction value A$_V$=1.4 
from \citep{schlafly2011measuringData} and the colour excess in the Gaia bands BP and RP E(BP-RP) =0.48 mag and 0.68 mag were derived.

We therefore used these values to estimate the distance modulus of (m-M)$_0$=14.12 mag and 14.31 mag corresponding to D=7.0 kpc and 7.3 kpc for FSR 19 and FSR 25 respectively. These values are very good in agreement with those obtained from the NIR photometry.

We have made use of theoretical isochrones to estimate the clusters metallicities and ages. Among the different models available, both  PARSEC  \citep{marigo2017new} and Dartmouth \citep{dotter2008dartmouth} isochrones seem to have gained 
acceptance for GC studies.

We present both set of models for NIR (VVVX) and optical (Gaia) CMDs. The best fits are shown in Figs.  \ref{Fig6}, \ref{Fig7}, \ref{Fig8} , \ref{Fig9}, for which we have used the derived distances and extinction values as explained above. Once these parameters were fixed, we overplotted  isochrones of different ages and then proceeded to varying the metallicity once an approximate age was derived. 
we note the close match to the datapoints that the derived ages and metallicities provide. These are 11 Gyr and [Fe/H]=-0.5 dex with $\alpha$-enhancement between 0 and $\pm$ 0.4 dex. No other combination could improve the 
agreement. Admittedly, isochrones fitting methods are usually a rough estimate prone to relatively large errors arising from distance and extinction determination and especially small numbers affecting the areas of interest.  In our case, this is also the case since the main-sequence turn-off is just below the limiting magnitude of the available photometry. In spite of this,  \ref{Fig6}, \ref{Fig7}, \ref{Fig8} , \ref{Fig9} all support the GC nature of the targets which is the main goal of this work.

Yet another parameter places FSR 19 and FSR 25 well in the regime of GC-candidates, namely their integrated luminosities.
For this purpose, we added the luminosities of all stars from the PM decontaminated diagrams using a GC size of 6$’$. Very likely this considers only RGB stars and neglects fainter, more numerous K and M dwarfs that could be cluster members.
For FSR 19 we obtained a total luminosity M$_{Ks}$ = -7.72 mag which translates to an absolute visual magnitude M$_V$ = -4.62 mag assuming a mean color $\big(V-Ks\big)$=3.1 mag given by \cite{bruzual2003stellar}  assuming the Chabrier initial mass function with age 11 Gyr and metallicity Z=0.02 for the cluster. Similarly, FSR 25 shows a total luminosity M$_{Ks}$ = -7.31 mag which corresponds to M$_V$ = -4.21 mag. We emphasize that these values should be considered as rough lower limits
for the luminosities. At a first glance, FSR 19 and FSR 25 could be close to the low-luminosity tail of the MW GCLF, $\sim$ 3 mag and 3.2 mag fainter than the peak of the MW GCLF (M$_V$= (-7.4$ \pm$0.2) mag from \citep{ashman1998globular,harris1991globular}.

Additionally, we searched for more evidence of an old stellar population that could be related to the targets. RR Lyrae are well know population II tracers and  are usually found in GCs. However, FSR 25 does not host a known RR Lyrae population and there is only  one potential candidate within the tidal radius of FSR 19. This is likely a fundamental mode pulsator or RRab  and has the Gaia ID: 4118273101762688128 and it is located at RA= 17:35:41.22; DEC= -21:02:16.9 (J2000). Its proper motion $\mu_{\alpha^\ast}$= 3.71$ \pm$ 0.21 mas $yr^{-1}$; $\mu_{\delta}$=-2.30 $\pm$ 0.16 mas $yr^{-1}$) is consistent (within 3 $\sigma$) with the mean  proper motion estimated for FSR 19. Its mean G  magnitude is  $17.517 \pm 0.025$ and color BP-RP=1.32. With a period P=0.4436972$\pm$ 0.0000003 days it is also visible in our VVVX photometry, with
J=15.42, H=15.97 and Ks=14.93 mag. The expected number of RRab in these fields relatively far from the centre is estimated to be about 55 RRab per square degree, which gives an expected frequency of 0.00009 RRab inside the tidal radius of FSR19 \citep{navarro2021rr}.

RR Lyrae stars are powerful distance indicators. We have adopted the period-luminosity relation of  \cite{muraveva2018rr} to estimate a distance to FSR 19 which turns out to be D=7.2$\pm$0.22 kpc. This value is in an excellent agreement with the distance measured from the other methods (see above) and provides strong evidence for the  RR Lyrae to be indeed a cluster member.

Finally, we have also computed the radial profiles for both FSR 19 and FSR 25 with the purpose of ascertaining the true physical sizes of these targets. We adopted the centres as documented in table \ref{table:1} 
for FSR 19 and  FSR 25.

\begin{figure*}
   \centering
   \includegraphics[width=\linewidth, height=6cm]{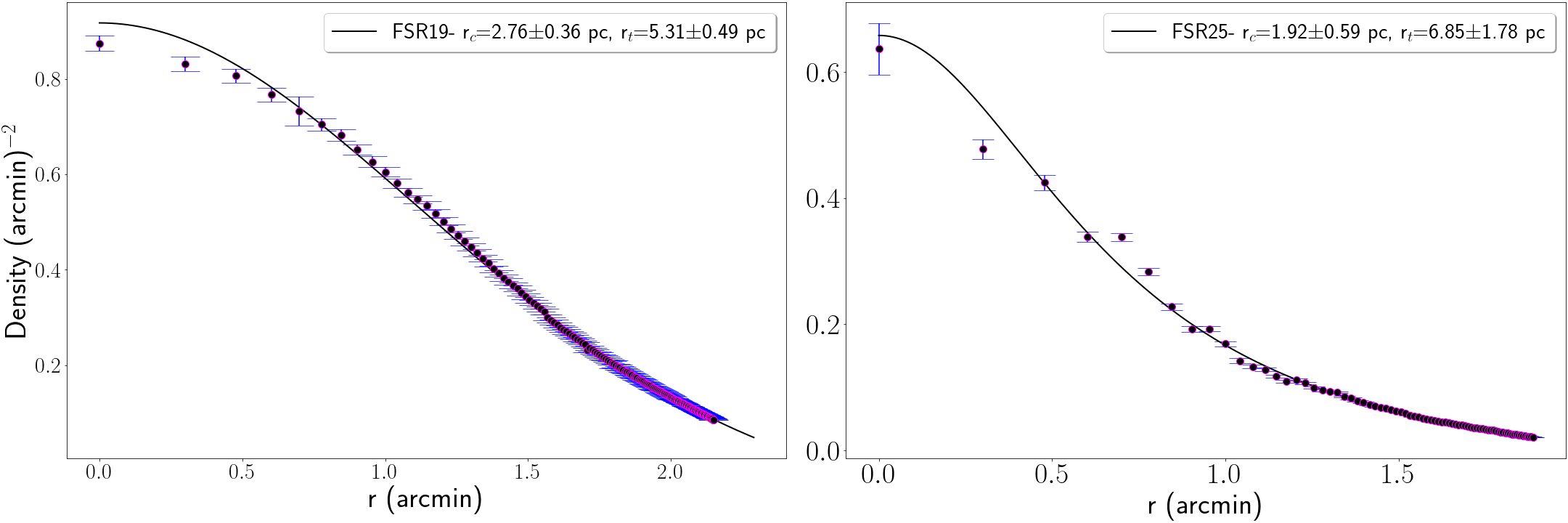}
   \caption{Radial profiles distribution of the FSR 19 and FSR 25 stellar densities as a function of radii fitted with \citep{king1962structure} profile. The black line represents the fit.}
              \label{Fig10}%
    \end{figure*}

We divided our sample into  circular annuli with increasing radii of 0.067$’$, out to a outer radius of 2$’$. This  provided an adequate balance between enough statistics  of  member stars and least contamination from interlopers (mostly foreground stars). Thereafter, the surface density was plotted as function of the mean distance to the circular annulus. The King profile \citep{king1962structure} is a long-used function to fit this density, even if the true shape is likely more complex. However for the purpose of establishing the GC nature and to compare the resulting parameters with those of similar compact systems, we provide the results using this approach. Figure   \ref{Fig10} shows that a King profile is indeed a good representation of the the projected number density. 
Models under-predicting the central density have been interpreted as an indication of GCs undergoing core collapse processes \citep[e.g][]{zocchi2016testing}. For our study, the innermost radial bins show a slight deficiency of member stars. Although this could hint to disruption processes, low number statistics and centre estimations are crucial and need to be improved before such claim can be supported. For FSR 19 we obtained $r_c$ = 1.30$'$ which at the adopted distance corresponds to 2.76$\pm$0.36 pc . Similarly, r$_t$ = 2.50$'$ which corresponds to 5.31$\pm$0.49 pc. For FSR 25, the resulting sizes were $r_c$ = 0.87$'$ (or 1.92$\pm$0.59 pc) and r$_t$ of 3.10$'$ (6.85$\pm$1.78 pc). These values are consistent with the typical MW GC radii listed in the 2010 \cite{harris1996catalog} compilation.

\section{Discussion} \label{sec:Dis}

As explained in Section \ref{sec:phyical}, both targets are located in the vicinity of filamentary dark nebulae and the complex extinction pattern in this region of the MW makes it important to consider if FSR 19 and FSR 25 could be just overdensities caused by an in-homogeneous differential extinction \citep[see e.g.,][]
{dutra2002low,koposov2007discovery,bica2011star,bidin2011three}, or even real clusters that are partially occulted by a dark cloud \citep[like the recently discovered giant GC FSR1758,][]{cantat2018gaia,barba2019sequoia}. Exploring the existing optical extinction maps \citep{schlegel1998maps,schlafly2011measuringData} we find no indication of reduced extinction in the fields of FSR 19 and FSR 25 that would suggest this possibility.

Fortunately, the multi-colour NIR photometry enables us to detect and map such windows of low extinction, that are interesting in their own because they allow us to pierce through the interstellar medium and probe the far sides of our Galaxy \citep[e.g.][]{dutra2002low,saito2020vvv,minniti2018new}. However, we find no such NIR window of low extinction to match the positions of these clusters, strengthening their genuine cluster nature.

It is interesting to note that \cite{Gran} discarded FSR 19 and FSR 25 as true GCs based on Gaia DR2 data. We note that the automatic procedure they adopt is sensitive to GCs that have distinct halo kinematics, but could in fact miss true GCs if their motions are similar to the MW bulge. We do not have radial velocities for our targets and VVVX PMs are not available for these clusters because the time baseline of observations is still too short $\sim$ (2 years). However, we have provided evidence for a number of physical tests which all point to the same direction: the presence of a real overdensity, the striking difference in the LF of their stellar populations compared to the field from a strong KS test, the appearance of the CMDs, the RR Lyra associated to FSR 19, the ages and metallicities, structural parameters and the Gaia PMs signatures are clear, and confirm the real  GC nature.

Thus, in the light of all these independent findings, it seems that both  FSR 19 and FSR 25 are genuine bulge GCs. They could very well  belong to the the few surviving  GCs that formed from the initial protogalactic collapse of the Galaxy. Many of these GCs have undergone disruptive processes. Our results are consistent with the works of \cite{kharchenko2013global} who measured the tidal radius of these new clusters to be within 5 to 10 pc. Again, \cite{kharchenko2016global} measured some properties of the FSR sample using 2MASS and obtained a distance of 5.98 kpc and absolute luminosity $M_{Ks}$ = -7.96 mag for FSR 19 which is in agreement with the measurements we obtained , a distance D = 7.2 $\pm$ 0.7 kpc and absolute luminosity $M_{Ks}$ = -7.72 mag, similarly for FSR 25 they obtained a distance of 6.67 kpc and an absolute luminosity $M_{Ks}$ = -8.80 mag while the distances agree within 3$\sigma$, the total luminosity is 1.5 mag brighter than ours  this could be an effect of the sample size used. The initial distance measurements done  by \cite{buckner2013properties} using the H band under-estimated  the actual distances  to the two targets by 3.4 kpc. In addition, we measured the ages, metallicities, core and tidal radii, distance  moduli and the absolute visual magnitude of the clusters as shown in table \ref{table:1}. Considering that we have relied only on the isochrones for estimating the ages of these clusters, we caution that the true ages of these clusters are uncertain, but in accordance with bulge globular clusters maybe placed between 10 and 14 Gyrs. FSR 19 and FSR 25 are both bulge GCs located about 8 deg. from the Galactic centre, and their derived integrated luminosities  show that these clusters are lying at the low end of the MW GC luminosity function. This explains partly why they were difficult to confirm in previous studies.

\begin{table}
\caption{The derived parameters of FSR 19 and FSR 25 Clusters}             
\label{table:1}      
\centering                          
\begin{tabular}{c c cr }        
\hline\hline                 
Parameter & FSR 19 & FSR 25 &  \\    
\hline                        
  $l$(deg) & 5.499 & 7.534 \\      
  $b$(deg) & 6.071 & 5.649   \\
  RA(J2000)(hh:mm:ss) &17:35:38.4 &17:41:43.2     \\
  DEC(J2000)(dd:mm:ss)& -21:04:12 &-19:34:16    \\
  $\mu_{\alpha^\ast}[$mas $yr^{-1}$] &-2.50 $\pm$0.76  &-2.61 $\pm$ 1.27\\
  $\mu_{\delta}$[mas $yr^{-1}$]& -5.02 $\pm$0.47 &-5.23 $\pm$0.74\\
  A$_{Ks}$[mag]& 0.19$\pm$0.07& 0.27$\pm$0.01  \\ 
  E(J-Ks)[mag]&0.38$\pm$0.06&0.54$\pm$0.09\\
  (m-M)$_0$[mag]&14.29$\pm0.08$&  14.23$\pm$0.06\\
  D [kpc]&7.2$\pm$0.7  & 7.0$\pm$0.9 \\
  M$_{Ks}$ [mag]& -7.72 &-7.31\\
   M$_V$ [mag]& -4.62 &-4.21\\
   $[Fe/H]$[dex]&-0.5 &-0.5\\
   Age [Gyr]& 11&11\\
   R$_c$ [pc]&2.76$\pm$0.36& 1.92$\pm$0.59\\
   R$_t$ [pc]&5.31$\pm$0.49 &6.85$\pm$1.78\\
\hline                                   
\end{tabular}
\end{table}


%
  

\section{Conclusions} \label{sec:concl}
The combination of the clean photometric catalog we obtained from the VVVX, Gaia EDR3 and 2MASS archives have allowed us to demonstrate that FSR 19 and FSR 25 are both genuine low luminosity GCs. We find for FSR 19  a reddening $E\big(J-K_s\big)$= 0.38$\pm$0.06  mag, and  extinction $A_{Ks}$= 0.19 $\pm$0.07 mag, and  measure a distance D= 7.2 $\pm$ 0.7 kpc, based on the mean magnitudes of the RC $K_s$ = 12.85 $\pm$ 0.05 mag.  We computed a total luminosity of $M_{Ks}$= -7.72 mag, and  measured its structural parameters to be  $r_c$= 2.76$\pm$0.36 pc and $r_t$= 5.31$\pm$0.49 pc. Based on the fit to theoretical isochrones we also estimate  a metallicity [Fe/H] = -0.5 dex, and an age t $\thicksim$ 11 Gyr for this cluster. 

For the GC FSR 25 we obtain a reddening $E\big(J-K_s\big)$= 0.54$\pm$ 0.09 mag, and  extinction $A_{Ks}$= 0.27 $\pm$0.01 mag, and  measure a distance D= 7.0$\pm$0.6 kpc, based on the mean magnitudes of the RC $K_s$= 12.88 $\pm$ 0.06 mag.  We computed a total luminosity of $M_{Ks}$= -7.31 mag, and measured its structural parameters to be  $r_c$= 1.92$\pm$0.59 pc  and $r_t$=6.85$\pm$1.78 pc. Based on the fit to theoretical isochrones we also estimate  a metallicity [Fe/H] = -0.5 dex, and an age t $\thicksim$11  Gyr for this cluster.

 Our newly confirmed clusters will indeed help increase the sample size of the MW GC system, especially in the innermost regions where they are most difficult to detect. Further work like spectroscopic observations are needed to compare the chemical signatures and orbital parameters for these new bulge globular clusters.
   
\begin{acknowledgements}
      We gratefully acknowledge the use of data from the ESO Public Survey program IDs 179.B-2002 and 198.B-2004 taken with the VISTA telescope and data products from the Cambridge Astronomical Survey Unit.
This publication makes use of data products from the Two Micron All Sky Survey, which is a joint project of the University of Massachusetts and the Infrared Processing and Analysis Center/California Institute of Technology, funded by the National Aeronautics and Space Administration and the National Science Foundation.
This work has made use of data from the European Space Agency (ESA) mission
{\it Gaia} (\url{https://www.cosmos.esa.int/gaia}), processed by the {\it Gaia}
Data Processing and Analysis Consortium (DPAC,
\url{https://www.cosmos.esa.int/web/gaia/dpac/consortium}). Funding for the DPAC
has been provided by national institutions, in particular the institutions
participating in the {\it Gaia} Multilateral Agreement. We also acknowledge the comments of the anonymous reviewer whose positive feedback helped to improve the quality of this paper.
 Support for the authors is provided by the BASAL Center for Astrophysics and Associated Technologies (CATA) through grant AFB 170002, and by Proyecto FONDECYT No. 1170121. J.A.-G. acknowledges support from Fondecyt Regular 1201490 and from ANID – Millennium Science Initiative Program – ICN12\_009 awarded to the Millennium Institute of Astrophysics MAS.

\end{acknowledgements}

%
%
   \bibliographystyle{aa} 
   \bibliography{Ref_1.bib} 

\end{document}